# Atomic layer control of metal-insulator behavior in oxide quantum wells integrated directly on silicon

Kamyar Ahmadi-Majlan[1], Tong-Jie Chen[2], Zheng Hui Lim[1], Patrick Conlin[1], Ricky Hensley[1], Dong Su[3], Hanghui Chen[4], Divine P. Kumah[2§], and Joseph H. Ngai[1*]

[1]Department of Physics, University of Texas-Arlington, Arlington, TX 76019, USA
[2]Department of Physics, North Carolina State University, Raleigh, NC 27695, USA
[3]Center for Functional Nanomaterials, Brookhaven National Laboratory, Upton, NY11973, USA
[4]Institute of Physics, New York University Shanghai, Pudong, Shanghai 200122, China

We present electrical and structural characterization of epitaxial $LaTiO_3$/$SrTiO_3$ quantum wells integrated directly on Si(100). The quantum wells exhibit metallic transport described by Fermi-liquid behavior. Carriers arise from both charge transfer from the $LaTiO_3$ to $SrTiO_3$ and oxygen vacancies in the latter. By reducing the thickness of the quantum wells, an enhancement in carrier-carrier scattering is observed, and insulating transport emerges. Consistent with a Mott-driven transition in bulk rare-earth titanates, the insulating behavior is described by activated transport, and the onset of insulating transport occurs near 1 electron per Ti occupation within the $SrTiO_3$ well. We also discuss the role that structure and gradients in strain may play in enhancing the carrier density. The manipulation of metal-insulator behavior in oxides grown directly on Si opens the pathway to harnessing strongly correlated phenomena in device technologies.

*jngai@uta.edu, §dpkumah@ncsu.edu

I. **INTRODUCTION**

Elucidating how the behavior of electronic materials evolves as material dimensions approach the nanoscale is of great fundamental and technological importance, especially as device technologies continue to shrink to smaller length scales. Transition metal oxides that exhibit strongly correlated phenomena are of particular interest, given their high carrier densities, short electronic length scales, and diverse range of phases they possess [1]. Recent advancements in thin film epitaxy have enabled correlated oxide films to be grown with thicknesses that approach fundamental electronic length scales (*e.g.*, Thomas-Fermi screening length). At such film thicknesses, altering the thickness by merely one or two unit-cells can have a profound effect on the stability of electronic phases [2,3]. Interfaces between oxides have also emerged as a setting in which high density electron gases can be created (*e.g.* $LaAlO_3/SrTiO_3$, $RETiO_3/SrTiO_3$ $RE =$ La, Gd, Nd *etc.*) [4-7]. Rich phenomena, including superconductivity and magnetism, have been observed in such electron gas systems [8-10]. In essence, interfaces in layered oxide heterostructures enable artificial correlated materials to be realized.

In this regard, artificial Mott systems exhibiting metal-insulator transitions that are driven by strong correlations are of particular interest for potential use in device applications [11]. Rare-earth titanates are an archetype example of a so-called filling-controlled Mott system, in which the metal-insulator transition is driven by carrier density. In bulk rare-earth titanates, carrier density is tuned through chemical composition [1,12,13]. In comparison, carrier density in artificial Mott systems can be tuned through confinement at interfaces. In particular, Moetakef et al. demonstrated that metal-insulator behavior in $GdTiO_3/SrTiO_3$ quantum wells could be achieved by reducing the thickness of the $SrTiO_3$ layer [6]. However, to achieve high carrier densities to induce a transition, $SrTiO_3$ thicknesses approaching 1 unit-cell were required. At such thicknesses carrier scattering from the substrate also becomes sizeable, which introduces a source of extrinsic disorder. Increasing the number of carriers in the system is necessary to realize a Mott driven transition in thicker quantum wells. In addition, device applications require integration on technologically relevant platforms, whereas growth on single crystal oxide substrates has been predominantly explored thus far [11,14,15].

Here we present the electrical and structural characterization of rare-earth oxide quantum wells integrated directly on Si(100) [16,17]. Our heterostructures are comprised of the Mott insulator LaTiO$_3$ (LTO) and band insulator SrTiO$_3$ (STO) [12]. At (100)-oriented heterojunctions between LTO and STO, an electron gas is created in the latter, due to a transfer of charge from the former [18-21]. To achieve metal-insulator behavior in thicker quantum wells, we further enhance sheet carrier densities by introducing oxygen vacancies in the STO. Nominally thick wells are metallic, exhibiting transport characteristics described by Fermi liquid behavior. We find that by reducing the thickness of the quantum wells, carrier-carrier scattering is enhanced, and a metal-insulator transition is induced at a well thickness of 6 unit-cells. Consistent with a Mott-driven transition in rare-earth titanates, activated transport over a large temperature regime is observed in the insulating phase, and the onset of insulating transport occurs as the carrier density within the quantum well approaches 1 electron per Ti. We also discuss the role that structure and gradients in strain may play in enhancing the carrier density. The control of metal-insulator transitions in oxides grown directly on Si opens the pathway to harnessing such strongly correlated phenomena in device technologies.

## II.    EXPERIMENTAL PROCEDURE

The LTO/STO heterostructures consist of 3 layers: a bottom layer of STO that is *n* unit-cells (u.c.) thick, an intermediary layer comprised of 3 u.c. of LTO and a top layer comprised of 1.5 u.c. of STO, as illustrated in the schematic of Fig. S1. The STO/LTO/Si heterostructures were grown using reactive MBE in a custom-built chamber operating at a base pressure of $< 2 \times 10^{-10}$ Torr. Undoped, 2" diameter Si (100)-oriented wafers (Virginia Semiconductor) were introduced directly into the MBE chamber. Prior to film deposition, the wafers were cleaned by exposing to activated oxygen, generated by a radio frequency source (VEECO) to remove residual organics from the surface. Two monolayers of Sr were deposited at a substrate temperature of 550 °C, which was subsequently heated to 870 °C to remove the native layer of SiO$_x$ through the formation and desorption of SrO [22]. A $2 \times 1$ reconstruction was observed in the reflection high energy electron diffraction (RHEED) pattern, indicating a reconstructed Si surface. Then a half monolayer of Sr was deposited at 660 °C to form a template for subsequent layers of STO. The substrate was then cooled to room temperature, at which 3 ML of SrO and 2 ML of TiO$_2$ were co-deposited at room temperature and then heated to 500 °C to form 2.5 unit-cells of crystalline STO. Subsequent layers of STO of various thicknesses were grown at substrate temperature of 500 °C. With the initial 1 monolayer of SrO at the interface, the STO films grown through co-deposition of SrO and TiO$_2$ are SrO, or A-site terminated. Immediately prior to depositing LTO, a single monolayer of TiO$_2$ was deposited to enable B-site terminated growth, which minimizes intermixing of Sr and La during subsequent deposition of 3 u.c. of LTO. Also, the STO was briefly annealed at 580 °C in ultra-high vacuum to introduce additional carriers through the formation of oxygen vacancies, immediately prior to deposition of LTO. All heterostructures were grown in an oxygen background pressure of $3 \times 10^{-7}$ Torr, which pressure avoids the formation of insulating phases such as La$_2$Ti$_2$O$_7$. Figure S2(a) shows typical post-growth RHEED images taken during growth along the [10] and [11] crystal directions. Finally, films were capped *in situ* at room temperature with 7 nm of amorphous silicon to protect the thin oxide layers.

Electrical transport of the heterostructures were measured in the van der Pauw geometry. Ohmic contacts were established on the four corners of 4 mm × 4 mm samples using Al wedge bonding. Resistivity and Hall measurements were performed using a Keithley 2400 sourcemeter and a Keithley 2700 multiplexer in a Quantum Design Physical Property Measurements System (PPMS).

Synchrotron X-ray diffraction measurements were measured at the 33ID beamline at the Advanced Photon Source at room temperature. Diffraction intensities were measured using a Pilatus 100K 2D detector with an incident photon wavelength of 0.799 Angstroms. Crystal truncation rods (CTRs) were measured along directions in reciprocal space defined by bulk Si.

## III.     RESULTS & DISCUSSION

Atomically abrupt interfaces between STO and Si and STO and LTO were achieved, as shown in the high angle annular dark-field (HAADF) image of the $n = 6$ quantum well in Fig. 1(a). Figure S2(b) shows synchrotron X-ray diffraction data (dots) of the $n = 5$, 6, 7, and 11 heterostructures taken along the (00L) specular direction as well as fits (solid lines) to the data. The finite thickness fringes that are clearly resolved also attest to the abrupt nature of the surfaces and interfaces of our quantum wells. Electrical transport measurements were performed on a series of heterostructures with different $n$, which ranged from $n = 4$ to 15 u.c. Hall measurements indicate that the sheet carrier densities $n_s$'s of the heterostructures are independent of $n$ (Fig. S2(c)), consistent with conductivity due to an electron gas at the interface between LTO and STO. We find that $n_s$'s for all the heterostructures are near $\sim 2 \times 10^{15}$ cm$^{-2}$, which is much higher than the $\sim 3.5 \times 10^{14}$ cm$^{-2}$ expected strictly from charge transfer from LTO to STO [6]. Carriers transferred from the LTO to the top 1.5 u.c. of STO are likely localized given the thinness of the STO, and proximity to the amorphous Si cap. We found that the thin STO cap was necessary to enable metallicity in the STO below [7]. Thus, our heterostructures can be thought of as quantum wells in which itinerant carriers are confined within a STO channel situated between LTO and Si.

A metal-insulator transition is observed in the transport characteristics of our quantum wells as the thickness of the STO $n$ decreases, as shown in Fig.1(b). Insulating samples are defined as those in which $dR_s/dT < 0$ throughout the temperature range of measurement. For reference, the 2D quantum of resistance (grey dashed line, Fig. 1(b)) is shown. The MI transition exhibits characteristics consistent with a transition that is driven by strong correlations, as opposed to disorder. First, Fermi liquid behavior is observed on the metallic side of the transition, which is characterized by a quadratic temperature dependence of $R_s$ given by $R_s = AT^2 + R_0$, as shown in Fig. 1(c). As the transition is approached with decreasing $n$, the $A$ parameter of the Fermi liquid $T^2$ dependence increases, as summarized in Fig. 1(d). The increase in $A$ corresponds to an increase in the effective mass of the carriers, which is a signature of enhanced electron-electron scattering in bulk rare-earth titanates [1,6,12,13,23]. Second, the insulating behavior in our heterostructures is described by Arrhenius, *i.e.* activated, transport as shown in the fit (red-dashed) in Fig. 1(e). In contrast, disorder driven insulating behavior is typically described by variable-range-hopping. We note that the temperature range over which Arrhenius behavior is observed is virtually identical to the range found in bulk LTO [12]. Third, the activation energies $E_A$'s extracted from fitting the insulating $n = 6$ and 5 quantum wells to the Arrhenius function are exceptionally small ($\sim 2$ and 3 meV, respectively), consistent with a system that is very close in proximity to a metal-insulator transition. Fourth, the slight increase in $E_A$ observed between the $n = 6$ and $n = 5$ quantum wells (inset of Fig. 1(e)) denotes a continued trend in enhanced correlations.

We now turn to the structural characterization of our LTO/STO quantum wells. Synchrotron X-ray diffraction reveals a rapid relaxation in epitaxial strain in the quantum wells. The basic perovskite lattice constants of bulk LTO, STO and that of the (100) surface of Si are 3.95 Å, 3.91 Å and 3.84 Å, respectively, thus, the LTO/STO heterostructures are, in part, under compressive strain. Figure 2(a) shows reciprocal space maps (RSM) for $n$ = 5, 6, 7, 11 samples around the off-axis reflection of Si (2, 2, L) for which there is no overlap with a Si Bragg peak. The RSMs are plotted in terms of the reciprocal lattice unit (r. l. u.) of bulk Si ($2\pi \cdot 5.431^{-1}$ Å$^{-1}$). Note the diffracted intensity centered at H = K = 2, L = 2.702 Si r.l.u. labelled C on the $n$ = 5 RSM in Fig. 2(a), denoting the presence of a component of the heterostructure that is coherently strained to the Si substrate. Less intense spectral weight is observed at lower H = K and larger L values on the $n$ = 5 RSM (labelled A and B), indicating the presence of partially relaxed components of the heterostructure and gradients in strain. As $n$ increases by just a single unit-cell, the relative intensity of the coherently strained (partially relaxed) component(s) C (A, B) decreases (increases), as shown more clearly in the line profile plots of Fig. 2(b). The extracted in-plane and out-of-plane lattice constants of components A, B and C are compared in Table S1 for the $n$ = 5, 6, 7 and 11 quantum wells. Based on the lattice constants extracted, we attribute components A and B to partially relaxed STO/LTO and STO respectively, while component C corresponds to the fraction of the STO that is coherently strained to the Si.

Film relaxation occurs via the formation of dislocations, which provides an extrinsic source of carrier scattering. However, carrier scattering from dislocations cannot explain the metal-insulator transition that we observe. We find that the thinner more strained (thicker more relaxed) heterostructures for which the density of dislocations should be fewer (greater) is insulating (metallic). To experimentally rule out extrinsic origins to the metal-insulator behavior, we have also studied ultrathin films of $La_{0.75}Sr_{0.25}TiO_3$ films grown on Si. We find that $La_{0.75}Sr_{0.25}TiO_3$ films of comparable thickness to the insulating quantum wells remain metallic, indicating that the emergence of insulating behavior is a characteristic of the quantum well, and not a generic effect of reducing the thickness of nominally metallic oxide films grown on Si (see Fig. S3).

To account for the enhancement of $A$ in the Fermi liquid transport and the emergence of insulating transport described by activated transport, we instead analyze the metal-insulator behavior within the context of strong correlations, which is a hallmark feature of rare-earth titanates. Reducing the thickness $n$ of the STO channel increases the 3-dimensional carrier density $n_{3D}$, which can lead to a filling-controlled Mott transition in bulk rare-earth titanates near carrier densities of 1 electron per Ti site, i.e., $n_{3D} \sim 1.6 \times 10^{22}$ cm$^{-3}$ [1,6,21,24].

To estimate $n_{3D}$, we model our quantum wells using coupled Poisson and Schrödinger equations [25]. The insulating $n = 6$ heterostructure for which the onset of insulating behavior occurs is modelled as a ~ 24 Å wide STO well situated between Si and LTO, in which the $z$-coordinate $z = 0$ ($z = -24$ Å) represents the STO/LTO (Si/STO) interface. The carrier density $n_{3D}$ is obtained from $n_{3D}(z) = \frac{m^*}{\pi\hbar^2}\sum_i(E_F - E_i)|\psi_i(z)|^2$ in which the sub-band energies $E_i$ and $\psi_i(z)$ are found by numerically diagonalizing the Schrödinger equation. We use $m^*$ of ~ $4m_e$ ($m_e$ = bare electron mass) which is appropriate for the Ti $t_{2g}$ bands of STO [26,27]. $E_F$ is determined by filling occupied sub-bands $i$ until $n_s \sim 2 \times 10^{15}$ cm$^{-2}$ is reached, which is the value of $n_s$ for our quantum wells. The carrier potential $V$ is determined by iteratively solving the Poisson $\nabla^2 V = -\rho_f(z)/\epsilon(E)$ and Schrödinger equations. $\rho_f(z)$ is comprised of $n_{3D}(z)$, positive ionized La ion cores situated at $z = 0$, as well oxygen vacancies in the STO, which we model using an exponentially decaying profile $\rho_{ox} \propto e^{z/\kappa}$ for which $z < 0$, and $\kappa = 8$ Å. A 0.2 V drop in potential at the interface between STO and Si is used to model the type-II band arrangement between the former and latter [28]. Finally, our model also accounts for the electric-field dependent dielectric constant of STO via $\epsilon(E(z)) = 1 + \epsilon_0^{-1}\partial P/\partial E$, where $P$ is related to $E(z)$ through the Landau-Ginzburg-Devonshire free energy [23]. Figure 3(a) shows the calculated $n_{3D}$, revealing that $n_{3D} \sim 1.6 \times 10^{22}$ cm$^{-3}$ can indeed be achieved given the $n_s$ and confinement of our quantum wells. These calculations provide further support for interpreting the metal-insulator behavior of our quantum wells within the context of a Mott transition.

We make a few remarks on the effect(s) that physical structure could have on the electronic behavior of the quantum wells. The gradient in strain in the STO channel could enhance $n_{3D}$ through an electric field $E_{STO}$ produced by the flexoelectric effect. To understand the origin of $E_{STO}$, first we note that the strained components of STO for the $n = 5$, 6, and 7 heterostructures exhibit anomalously large $c/a$ ratios that indicate that the volume of the STO unit-cell is not conserved according to the predictions of elastic theory. Prior studies revealed that this non-conservation in unit-cell volume arises from non-centrosymmetric displacements of the Ti cations, which gives rise to a large upward polarization, as illustrated in the inset of Fig. 3(b) [29-31]. As the quantum well becomes thinner, the polarization of the strained component of STO (component C) increases, as evidenced by the $c/a$ ratios shown in Fig. 3(b) for the $n = 5$, 6, 7 and 11 quantum wells. The horizontal dashed line in Fig. 3(b) indicates the $c/a$ ratio above (below) which strained STO is polar (non-polar). We note that *ab initio* density functional theory (DFT) calculations also predict the upwards polarization in STO even with the presence of LTO on top, as summarized in Fig. S4. The coherently strained component of STO, however, is only ~ 2 - 3 u.c. thick, as the compressive strain decays very rapidly towards the LTO. Figure 3(c) illustrates the relaxation of the coherently strained component of STO for the *n* = 5 quantum well. Analysis of the RSM indicates that within ~ 2 u.c. the extracted in-plane lattice *a* increases from 3.84 Å to 3.89 Å, *i.e.* essentially bulk STO (see Table S1). A key consequence of the rapid relaxation is the creation of a large gradient in strain. Such a gradient in strain can, in principle, generate an electric field via the flexoelectric effect [32,33]. Based on the rapid relaxation of the STO, we estimate $E_{STO}$ could be as large as ~ $8 \times 10^7$ V/m using a phenomenological approach that is independent of the actual mechanism(s), *e.g.* dislocations, etc., for film relaxation (see Supplemental Information) [33].

The polar distortions in the strained STO could also affect carrier bandwidth. In principle, the presence of polar distortions in the STO should decrease overlap between Ti 3d and O 2p orbitals, as the Ti-O-Ti bond angle between adjacent unit-cells is perturbed from 180°. A decrease in overlap between Ti 3d and O 2p orbitals would weaken carrier bandwidth, thereby promoting the emergence of an insulating state [2]. However, STO grown on Si experiences compressive strain, which enhances orbital overlap by reducing the Ti-O bond length. Consequently, analysis indicates that a reduction in bond length in combination with a reduction in bond angle actually modestly enhances the carrier bandwidth (see Supplemental Information).

Finally, for a Mott driven transition with 1 electron per Ti, accompanying signatures of charge and or spin ordering are expected to arise [1]. Synchrotron XRD was used to search for superstructures in the periodicity of the lattice associated with spin or charge ordered states. At present, conclusive evidence for superstructures remains elusive, though we note that the thinness of the STO renders detection of minute structural distortions challenging. Furthermore, we again note that the activation energies of the Arrhenius transport we observe are exceptionally small (2 - 3 meV), indicating that the STO is "barely" in the insulating regime.

### IV. SUMMARY

In summary, we have demonstrated atomic-scale control of strong correlations and metal-insulator behavior in LTO/STO quantum wells integrated directly on Si(100). Consistent with a Mott driven effect, the insulating behavior is described by activated transport and the metal-insulator transition occurs as the carrier density of STO approaches 1 electron per Ti. Oxide heterostructures in which strong correlations can be tuned through thickness could be exploited as channel materials in emerging field-effect devices, or be exploited in sensors or possibly energy harvesting [34-36]. The material behavior of strongly correlated oxides complements the properties of conventional semiconductors, and could lead to entirely new modalities in device functionality.

**ACKNOWLEDGMENTS**

This work was supported by the National Science Foundation (NSF) under award No. DMR-1508530 and the University of Texas-Arlington under the Research Enhancement Program. We also thank the Alan G. MacDiarmid Nano Tech Institute at the University of Texas-Dallas for use of the Quantum Design PPMS. STEM work is supported by the Center for Functional Nanomaterials, Brookhaven National Laboratory, which is supported by the U.S. Department of Energy (DOE), Office of Basic Energy Science, under Contract No. DE- SC0012704. We express our gratitude to S. Ismail-Beigi for providing the code for the Poisson-Schrödinger solver.

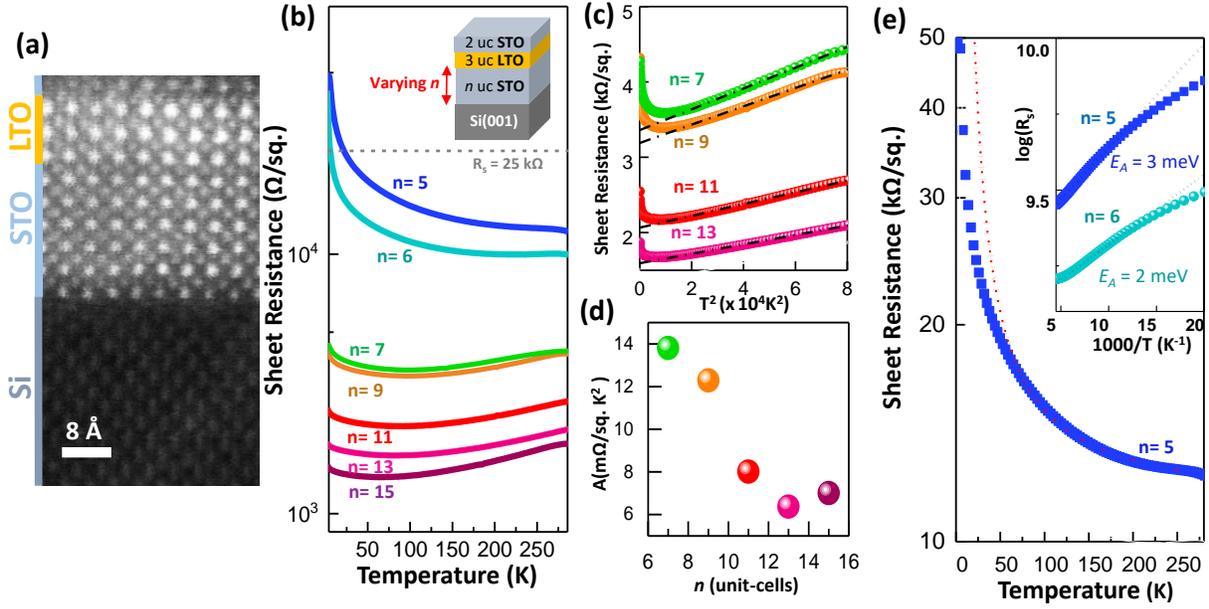

FIG. 1. (a) STEM HAADF image of the $n = 6$ quantum well. (b) Sheet resistance of LTO/STO/Si heterostructures of different thicknesses showing metal-insulator transition. The 2D Mott minimum conductivity is indicated by the dashed line. (c) Sheet resistance for metallic LTO/STO/Si heterostructures plotted versus $T^2$. The dashed lines are fits to the Fermi-liquid equation. (d) Temperature coefficient $A$ as a function of thickness for metallic samples. (e) Sheet resistance for insulating $n = 5$ heterostructure. The red dotted line is a fit to Arrhenius law. The inset shows sheet resistance plotted against $1/T$ for the insulating $n = 5$ and 6 heterostructures. Black dotted lines are fits to the activated transport, with activation energies shown.

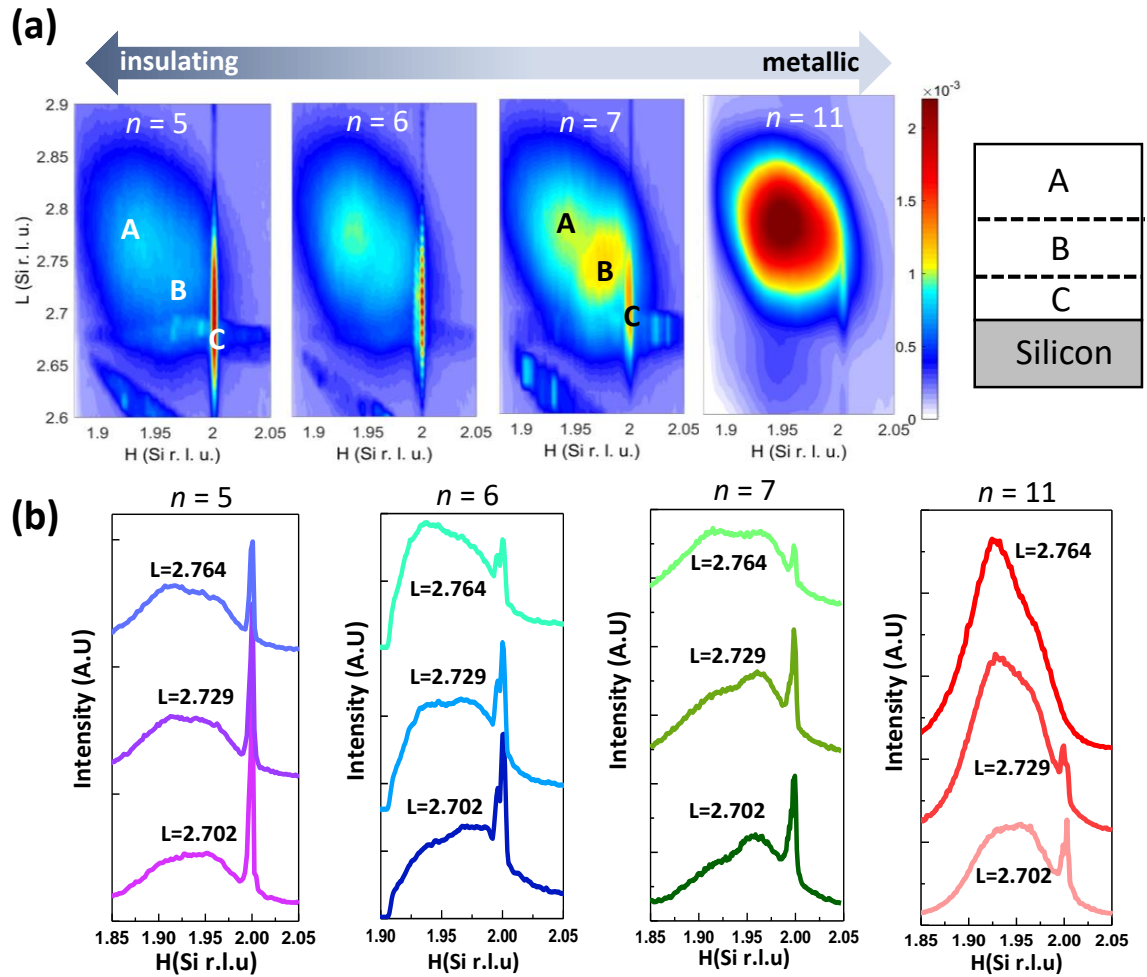

FIG. 2. (a) Reciprocal space maps taken of quantum wells of various thickness *n*, showing the partially relaxed (labelled A, B) and coherently strained regions (labelled C). (b) Line profile plots of the RSMs shown in (a).

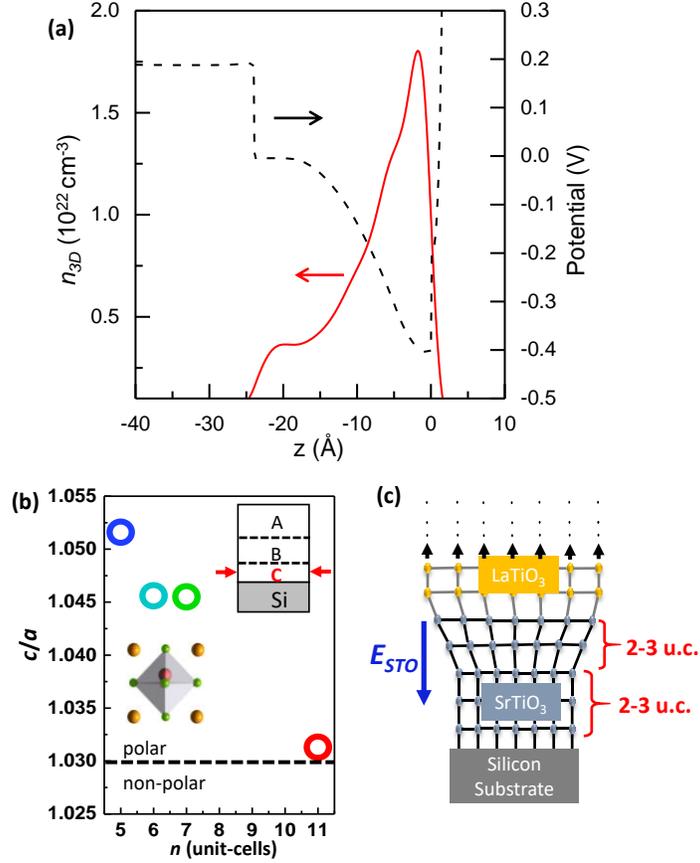

FIG. 3. (a) $n_{3D}$ calculated from a Poisson-Schrödinger model showing carrier density equivalent to Ti occupation of half-filling. (b) Ratio between $c$ to $a$ lattice constants for the component of STO ('C') that is coherently strained to Si (top inset). The anomalously large $c/a$ ratios indicate the STO exhibits an out-of-plane polarization, illustrated in the bottom inset. (c) Schematic illustrating the thickness of the coherently strained STO and the number of unit-cells over which strain relaxes for the $n = 5$ heterostructure. The gradient in strain produces an electric field via the flexoelectric effect.